\documentclass{mem}
\usepackage{natbib}\usepackage{txfonts}\usepackage{balance}
\usepackage{graphicx}
\usepackage{multirow}
\usepackage[a4paper]{hyperref}
\idline{81}{1}
\begin{document}
\def\teff{$T\rm_{eff }$}
\def\kms{$\mathrm {km s}^{-1}$}

\title{
Quasi-Stationary and Transient Patterns in Jets
}

   \subtitle{}

\author{
A.P. \, Lobanov
          }

  \offprints{A.P. Lobanov}

\institute{
Max-Planck-Institut f\"ur Radioastronomie, Auf dem H\"ugel 69,
53121 Bonn, Germany\\
\email{alobanov@mpifr.de}
}

\authorrunning{Lobanov}

\titlerunning{Stationary and Transient Jet Patterns}

\abstract{Apparent evolution of relativistic flows as traced by radio
emission results from a combination of several factors related to
propagation of relativistic blobs or shocks, velocity, density and
pressure stratification of the underlying flow, plasma instability and
(possibly also) phase and time travel effect. This combination can
create an intricate and chaotic patterns of the observed morphological
changes in radio emission, which complicates the analysis and
interpretation of kinematic and physical properties of the jet plasma.
Recent studies have indicated that slow and quasi-stationary patterns
in jets are most likely formed by plasma instabilities while faster,
superluminally moving patterns are related to highly relativistic
plasma condensations produced by the nuclear flares. Some of the
stationary patterns may also be related to recollimation shocks or
locations where strong non-thermal continuum is produced in jets.
Similarities and differences of the AGN and XRB jets in this respect
are reviewed.   
\keywords{ galaxies: jets -- stars: binaries: general -- stars: winds: outflows} } 
\maketitle{}

\section{Introduction}

Recent years have witnessed an increasingly wider recognition of the
ubiquity of relativistic outflows (jets) in galactic nuclei
\citep{lobanov_zensus1997,lobanov_falcke2001} and X-ray binaries
\citep{lobanov_remillard2006,lobanov_mirabel2010}.  This has turned jets
into an effective probe of the nuclear regions in external galaxies
\citep{lobanov_lobanov2006} and the physics of black hole -- accretion
disk systems in active galactic nuclei
\citep{lobanov_lobanov2008,lobanov_lobanov2010} and X-ray binaries
\citep{lobanov_mirabel2010}.

In the radio regime, very long baseline interferometry (VLBI) enables
direct imaging of spatial scales comparable the gravitational radius,
$R_\mathrm{g} = G\,M_\mathrm{bh}/c^2$, of the central black hole in AGN
using ground VLBI observations at 86\,GHz and higher
\citep{lobanov_krichbaum2008} and space VLBI observations at centimetre
wavelengths \citep{lobanov_takahashi2004}. Such
high-resolution radio observations also access directly the regions
where the jets are formed \citep{lobanov_junor1999}), and
trace their evolution and interaction with the nuclear environment
\citep{lobanov_lobanov2007,lobanov_lobanov2008,lobanov_middelberg2008}.

Jets in both AGN and XRB are formed in the immediate vicinity of the
central black hole, in the extreme
relativistic environment requiring a full general relativistic
magnetohydrodynamic (GRMHD) description for obtaining physically viable
models.  This overarching closeness of the basic physical settings has
prompted a number of investigations connecting together the observed
properties of XRB and AGN via their black hole masses, and radio and
X-ray luminosities (the so-called ``fundamental plane of accreting black
holes; {\em cf.,} \citeauthor{lobanov_merloni2003}
\citeyear{lobanov_merloni2003}, \citeauthor{lobanov_falcke2004}
\citeyear{lobanov_falcke2004}, \citeauthor{lobanov_koerding2008}
\citeyear{lobanov_koerding2008}), as well as via the characteristic
timescales of X-ray variability
\citep{lobanov_mchardy2006,lobanov_koerding2007}. Different X-ray
emission/hardness states observed in the XRB have also been suggested to
relate with different states of accretion and jet production
\citep{lobanov_mcclintock2003,lobanov_fender2004}.

Despite the overwhelming similarities, there is a number of aspects at
which XRB and AGN jets are different, and these aspects range from the truly
physical (for instance, hotter accretion disks and companion-induced
disk truncation in XRB, central neutron star (as opposed to black hole)
in low-mass XRB, barycentric motion, and strong precession induced by
the secondary star in XRB), to observational (XRB jets are typically
detected at a much lower dynamic range than their AGN counterparts), and
even to ``sociological'' ones.

The latter aspects arise sometimes from the AGN and the XRB communities
using somewhat divergent ``slangs'', which may lead to misunderstanding
and misconceptions. For instance, the usual comparison of superluminal
motions in XRB and AGN relates actually two quite different phenomena:
the superluminal features are detected in XRB at about 3--4 orders of
magnitude large distances (in terms of the gravitational radii of the
central black hole) than their AGN ``counterparts'', and they, in fact,
correspond rather to relativistic motions detected in the hotspots
of AGN lobes.

This paper provides a basic overview of the similarities and differences
in the observed manifestations of XRB and AGN jets. The discussion
is started with a brief review of the current understanding of the
physics of AGN jets, which is used as a basis for comparison with the
results obtained for XRB jets.

\begin{table*}[ht!]
\caption{Spatial and temporal scales in AGN and XRB jets}
\label{tb:regions}
\begin{center}
\begin{tabular}{l|c|c|c|c|c|c|c|c}
\hline\hline
\multirow{3}{*}{Region}        & \multicolumn{4}{|c|}{Spatial Scales} & \multicolumn{4}{c}{Temporal/Frequency Scales} \\\cline{2-9} 
& \multicolumn{2}{|c|}{Natural Scale [$R_\mathrm{g}$]} & \multicolumn{2}{c|}{Angular Scale} & \multicolumn{2}{c}{$10^8 M_\odot$ AGN} & \multicolumn{2}{|c}{$10 M_\odot$ XRB} \\\cline{2-9}
       & $l_\mathrm{z}$ & $l_\mathrm{r}$ & $\theta_\mathrm{z}$ & $\theta_\mathrm{r}$ & $\tau_\mathrm{z}$ & $\tau_\mathrm{r}$ & $\tau_\mathrm{z}$ & $\tau_\mathrm{r}$  \\\hline
Launching    & $10^2$     & $10^1$            & $0.1\,\mu$as         & $0.01\,\mu$as          & $15$\,hr  & $2$\,hr & $200$\,Hz  & $2000$\,Hz \\
Collimation  & $10^3$     & $3\times 10^2$    & $1\,\mu$as           & $0.3\,\mu$as          & $6$\,d  &  $2$\,d & $20$\,Hz  &  $60$\,Hz \\
Acceleration & $10^6$     & $5\times 10^4$    & $1$\,mas             & $0.05$\,mas           & $15$\,yr  &  $1$\,yr & $1$\,min  &  $2$\,sec \\
KFD Flow (Jet)& $10^9$    & $5\times 10^7$    & $1^{\prime\prime}$   & $0.05^{\prime\prime}$ & $15$\,kyr  &  $800$\,yr & $15$\,hr  &  $1$\,hr \\
Hotspot      & $10^{11}$  & $5\times 10^9$    & $100^{\prime\prime}$ & $5^{\prime\prime}$    & $15$\,Myr  &  $80$\,kyr & $60$\,d  &  $3$\,d \\
Lobe         & $10^{11}$  & $3\times 10^{10}$ & $100^{\prime\prime}$ & $30^{\prime\prime}$   & $15$\,Myr  &  $500$\,kyr & $60$\,d  &  $20$\,d \\
\hline
\end{tabular}
\end{center}
{\footnotesize {\bf Notes:}~Angular scales are calculated for a
$10^8M_\odot$ AGN at a distance of 1\,Gpc and for a $10\,M_\odot$ XRB at
a distance of 100\,pc. The index $z$'' indicates length scale along the
jet, and the index ``$r$'' indicates the length scale across the jet
(calculated assuming collimation by the magnetic field, MHD
acceleration, and adiabatic expansion of the flow. The corresponding
temporal/frequency scales pertain to the light crossing times of the
respective length scales.}
\end{table*}

\section{Basic physics of relativistic flows} 

Emission properties, dynamics, and evolution of cosmic relativistic jets
are intimately connected to the characteristics of the supermassive
black hole, accretion disk and (in the case of AGN jets) broad-line
region in the nucleus of the host galaxy \citep{lobanov_lobanov2008}.
The jet continuum emission is dominated by non-thermal synchrotron and
inverse-Compton radiation \citep{lobanov_unwin1997}. The synchrotron
mechanism plays a more prominent role in the radio domain, and the
properties of the emitting material can be assessed using the turnover
point in the synchrotron spectrum
\citep{lobanov_lobanov1998b,lobanov_lobanov1999}, synchrotron
self-absorption \citep{lobanov_lobanov1998a}, and free-free absorption
in the ambient plasma \citep{lobanov_walker2000,lobanov_kadler2004}.

Jets can be ``dissected'' (both qualitatively and quantitatively) into
five distinct regions, following the physical transitions and evolution
undergone by a relativistic flow along the way of its propagation: {\em
1)}~launching region; {\em 2)}~collimation region; {\em 3)}~acceleration
region; {\em 4)}~kinetic flux-dominated (KFD) flow; and {\em
5)}~dissipation (hotspot/lobe) region.  Spatial and temporal scales
related to these regions are presented in Table~\ref{tb:regions} for a
typical AGN and an XRB object. A brief discussion of each of these
regions follows below (cf., \citeauthor{lobanov_lobanov2010}
\citeyear{lobanov_lobanov2010} for a more detailed description).

\subsection{Jet Launching}

Jets in AGN and XRB should be formed in the immediate vicinity of the
central black hole (Camenzind \cite{lobanov_camenzind2005}), at
distances of $10$--$10^2\,R_\mathrm{g}$
\citep{lobanov_meier2001b,lobanov_meier2009}. The jets carry away a
fraction of the angular momentum and energy stored in the accretion flow
\citep{lobanov_blandford1982,lobanov_hujeirat2003} or corona
\citep{lobanov_merloni2002} and in the rotation of the central black
hole
\citep{lobanov_blandford1977,lobanov_koide2002,lobanov_semenov2004,lobanov_komissarov2005}.

The production of highly-relativistic outflows requires a large fraction
of the energy to be converted to Poynting flux in the very central
region \citep{lobanov_sikora2005}.  The efficiency of this process may
depend on the spin of the central black hole
\citep{lobanov_meier1999,lobanov_meier2001}. The collimation of such a
jet requires either a large scale poloidal magnetic field threading the
disk \citep{lobanov_spruit1997}) or a slower and more massive MHD
outflow launched at larger disk radii by centrifugal forces
\citep{lobanov_bogovalov2005}.  The flowing plasma is likely to be
dominated by electron-positron pairs
\citep{lobanov_wardle1998,lobanov_hirotani2005} although a dynamically
significant proton component cannot be completely ruled out at the
moment \citep{lobanov_celotti1993}.

At present, the scales of $\le 10^2\,R_\mathrm{g}$ are barely accessible
for direct imaging both in XRB and AGN, with notable exceptions of M\,87
($R_\mathrm{g}\approx 2\,\mu as$) and Sgr~A* ($R_\mathrm{g}\approx
5\,\mu as$). This limits currently observable manifestations of the jet
launching region to the domains of time variability and integrated
spectrum. Curiously enough, the transverse time scales (light crossing
time across the jet) in the launching region are rather similar to those
inferred from the high-frequency QPOs in XRB and intraday variability in
AGN. In both cases, other explanations (orbital motion in accretion disk
in XRB and scintillations in AGN) are often preferred, but the jets may
nevertheless play a role in producing this variability.

\subsection{Collimation Region}

Observations of jets in nearby AGN
\citep{lobanov_junor1999,lobanov_bach2005} indicate that the flows are
collimated on scales of $\sim 10^3\,R_\mathrm{g}$, thus indicating a
very efficient collimation mechanism. A number of recent works
(cf. \citeauthor{lobanov_meier2001} \citeyear{lobanov_meier2001},
\citeauthor{lobanov_koide2002} \citeyear{lobanov_koide2002},
\citeauthor{lobanov_gracia2005} \citeyear{lobanov_gracia2005},
\citeauthor{lobanov_komissarov2007} \citeyear{lobanov_komissarov2007},
\citeauthor{lobanov_porth2010} \citeyear{lobanov_porth2010}) have
demonstrated the paramount importance of magnetic field for efficient
collimation. The resulting collimation is achieved either via
self-collimation of the fast, relativistic part of the flow or via
collimation of the overdense and slower outer layers of the flow,
leaving a very narrow channel for propagation of the underdense
relativistic beam.

For the collimation region, the natural time scales (see
Table~\ref{tb:regions} for the respective light crossing times) are
close to the typical IDV timescales in AGN and low frequency QPO
timescales in XRB, suggesting that the jet plasma on these scales may
also contirbute to the respective variability.

\subsection{Acceleration Region}

Acceleration of the flow may be complete within about
$10^3\,R_\mathrm{g}$ \citep{lobanov_meier2009} or continue all the way
to scales of $\sim 10^6\,R_\mathrm{g}$ \citep{lobanov_vlahakis2004},
with growing amount of evidence favouring the latter scenario
\citep{lobanov_bach2005,lobanov_lee2008}.

At distances of $\sim$\,$10^3\,R_\mathrm{g}$, the jets become visible
in the radio regime (this region is often branded as ``core'' of the
jet). Recent studies indicate that at $10^3$--$10^5\,R_\mathrm{g}$
($\le\,1$\,pc, in AGN) the jets are likely to be dominated by pure
electromagnetic processes such as Poynting flux
\citep{lobanov_sikora2005} or have both MHD (kinetic flux) and
electrodynamic components \citep{lobanov_meier2003}.  At larger
scales, the jets are believed to be kinetic flux-dominated, while the
transition from Poynting flux-dominated (PFD) to kinetic
flux-dominated (KFD) flow remains elusive for the present day
observations. One possibility is that the transition occurs still in
the region which is optically thick at radio
wavelengths. Alternatively, the PFD flow may be sub- or even
non-luminous at all. In any case, it would be marginally resolvable
with VLBI, and would most likely be associated with the optically
thick ``cores'' in most AGN jets \citep{lobanov_lobanov1998a}, again
with notable exceptions of the nearest ones, such as M\,87
\citep{lobanov_junor1999}.

The
non-thermal continuum radio emission from the jet core does not appear
to have strong shocks, and its evolution and variability can be
explained by smooth changes in particle density of the flowing plasma,
associated with the nuclear flares in the central engine
\citep{lobanov_lobanov1999}.

This supports the suggestion that jets at these scales are
particle-dominated and they reach the equipartition regime downstream,
at large distances.
\citep{lobanov_unwin1997,lobanov_lobanov1998a,lobanov_hirotani2005}. Combining
these calculations with estimates of the jet kinetic power provides
strong indications that the relativistic fraction of the outflowing
material is most likely represented by the electron-positron plasma
\citep{lobanov_reynolds1996,lobanov_hirotani2000,lobanov_hirotani2005}.

\subsection{Kinetic Flux-Dominated (KFD) Flow}

The KFD flows, observed in AGN at scales of
$\sim 10^6$--$10^9\,R_\mathrm{g}$, provide bulk of the observational
material available for relativistic jets.  
Two distinct regions,
with different physical mechanisms dominating the observed properties,
can be identified in the KFD flows (cf.,
\citeauthor{lobanov_lobanov2010} \citeyear{lobanov_lobanov2010}):
1)~parsec-scale flows ($\sim$\,10\,pc scales) dominated by
relativistic shocks and 2)~larger-scale jets ($\ge$\,100\,pc) where
plasma instability gradually becomes dominant.  

This picture may be further complicated by transverse stratification
of the flow, with the jet velocity, particle density and magnetic
field changing substantially from the axis of the jet to its outer
layers (cf., \citeauthor{lobanov_aloy2001}
\citeyear{lobanov_aloy2001}, \citeauthor{lobanov_gomez2008}
\citeyear{lobanov_gomez2008}).  As a consequence of this
stratification, shocks and instabilities may in fact co-exist on all
scales in the jets \citep{lobanov_lobanov1998c,lobanov_lobanov2001},
with instabilities simply remaining undetected in compact flows, owing
to limited resolution and dynamic range of VLBI observations.

Parsec-scale outflows are characterised by
pronounced curvature of trajectories of superluminal components
\citep{lobanov_kellermann2004,lobanov_lobanov1999}) and rapid changes
of velocity and flux density \citep{lobanov_lister2009}.  The
superluminal components are often found to thread through an
underlying pattern consisting of one or several stationary features
\citep{lobanov_kellermann2004,lobanov_savolainen2006,lobanov_lister2009}.

Specific geometric conditions and extremely small viewing angles could
lead to formation of stationary features in relativistic flows
\citep{lobanov_alberdi2000}.  However, a more general and physically
plausible explanation is offered by standing shocks \citep{lobanov_daly1988,lobanov_gomez1995,lobanov_perucho2007a} or Kelvin-Helmholtz instability patterns
propagating at slow, subluminal speed
\citep{lobanov_hardee2000,lobanov_hardee2007,lobanov_lobanov2001}.

Stationary features observed near the jet base (at $\sim
10^6\,R_\mathrm{g}$ may indeed represent standing shocks and play a
major role in accelerating particles
(\citep{lobanov_becker2008,lobanov_arshakian2010,lobanov_leon2010}. Such
shocks could be responsible for the persistent high levels of
polarization in blazars
\citep{lobanov_darcangelo2007,lobanov_marscher2008}.  More
speculatively, these shocks could also be the sites of continuum
emission release due to conversion from Poynting flux-dominated to
kinetic flux-dominated flow.

Complex evolution of the moving shocked regions is revealed in
observations
\citep{lobanov_lobanov1999,lobanov_gomez2001,lobanov_jorstad2005} and
numerical simulations
\citep{lobanov_agudo2001,lobanov_mimica2007,lobanov_mimica2009} of
parsec-scale outflows. However, the shocks are shown to dissipate
rapidly \citep{lobanov_lobanov1999}, and shock dominated regions are
not likely to extend beyond $\sim$\,100\,pc. Starting from these
scales, the underlying flow shaped by Kelvin-Helmholtz instability
\citep{lobanov_hardee2000,lobanov_hardee2007} determine at large the
observed structure and dynamics of extragalactic jets
\citep{lobanov_lobanov1998c,lobanov_walker2001,lobanov_lobanov2001,lobanov_lobanov2003,lobanov_hardee2005,lobanov_perucho2006}.

\subsection{Hotspots and Lobes}

Jets are expected to gradually lose their collimation and expand into
large-scale lobes at distances of $\ge 10^9\,R_\mathrm{g}$. Appearance
of the hotspots (still expected to contain a highly-relativistic
plasma; \citeauthor{lobanov_georganopoulos2004}
\citeyear{lobanov_georganopoulos2004}) inside the large-scale lobes
suggests that this process can be gradual, affecting first outer
layers of the flow and working its way through the jet interior.  At
the same time, at scales exceeding $\sim 10^{10}\,R_\mathrm{g}$, the
helical surface mode of Kelvin-Helmholtz instability is likely to be
another factor working on disruption and destruction of the outflows
\citep{lobanov_lobanov2003,lobanov_lobanov2006,lobanov_perucho2007}.

It should be noted that most of the proper motion measurements made in
XRB refer in fact to the scales of $10^{10}$--$10^{12}\,R_\mathrm{g}$,
({\em cf.,} \citeauthor{lobanov_mirabel1994}
\citeyear{lobanov_mirabel1994}; \citeauthor{lobanov_corbel2002}
\citeyear{lobanov_corbel2002}), thus the motions detected in XRBs are
more likely to be related to the advance of the lobes and hotspots in
AGN \citep{lobanov_fomalont2001}. However, the latter have much
smaller advance speeds, which raises the question whether this
comparison can still be physically valid.

\section{Similarities and differences}

The above example of comparing spatial scales in XRB and AGN
underlines the general issue of relating the jets in these two object
classes. The overarching similarity of the general mechanism of matter
accretion powering bipolar outflows (outlined overwhelmingly by the
unification of the energy output in XRB and AGN made in the
fundamental black hole activity plane;
\citeauthor{lobanov_merloni2003} \citeyear{lobanov_merloni2003},
\citeauthor{lobanov_falcke2004} \citeyear{lobanov_falcke2004},
\citeauthor{lobanov_koerding2008} \citeyear{lobanov_koerding2008})
leads however to quite divergent paths in AGN and XRB outflows. This
divergence begins with the XRB accretion disks being hotter and
naturally truncated due to a presence of a companion star. It
continues then further to the composition of the flow and the
environment (with more heavy ions expected in XRB), differences in the
extended environment (potentially strong winds from the companion star,
and differences between the interstellar and intergalactic
environment), and dynamic differences (strong precession and
barycentric motion induced by the companion star). All these make the
comparisons and relations between XRB and AGN jets rather non-trivial.

One can start this discussion by bringing together the general
definitions and jargons used in the two fields, and then considering
the main areas in which the comparisons are usually made: the
temporary (variability) and spatial (morphology and dynamics) scales
in XRB and AGN.

In the XRB ``slang'', physical states of the outflow are often
categorised between ``steady, compact jets'' with hard (flat) spectrum
and ``transient, large scale jets'' with soft (steep) spectrum. The
AGN community uses overwhelmingly the spectral index as the main
``divider'', thus distinguishing between flat-spectrum emission from
ultra-compact jets (``cores'') and shocked regions and steeper
spectrum emission from the underlying flow, dying shocks and
instability-dominated regions. Thus, the steady XRB jets can be in
principle physically related to the flat-spectrum parts of AGN flows
(especially given that the shock dissipation scales of $\sim
10^8\,R_\mathrm{g}$ are normally unresolved in XRB). The transient XRB
jets are then probably more related to the steep-spectrum parts of AGN
jets and even lobes.

This picture has however several outstanding issues related to spatial
and especially temporal scales inferred for the respective
counterparts in AGN and XRB outflows.

\subsection{Spatial scales}

GRS 1915+105 features proudly a 100 AU-long ``compact, steady'' jet in
low-hard state \citep{lobanov_dhawan2007}. This jet is $\approx
10^9\,R_\mathrm{g}$(!!) long, and no AGN has ever been seen 
shows a flat spectrum radio emission of such an extent. Thus a simpleminded
connection between the ``low-hard'' XRB and ``flat-spectrum'' core of AGN (even a ``naked'' core, without a strong jet) may be problematic.

In the transient, large scale jet observed during the ``high-soft''
state of XRB, superluminal ejections are traced to separations of
$\approx 800$ mas (GRS 1915+105; \citeauthor{lobanov_mirabel1994}
\citeyear{lobanov_mirabel1994}) and 30$^{\prime\prime}$ (J1550-564,
\citeauthor{lobanov_corbel2002} \citeyear{lobanov_corbel2002}). These
separations correspond to scales of
$10^{10}$--$10^{12}\,R_\mathrm{g}$, which well into the scales of
radio lobes in AGN. Superluminal components in AGN jets, to which the
XRB jets are commonly compared, are typically detected at
$10^4$--$10^7\,R_\mathrm{g}$, and therefore the physical conditions in
theses superluminal features should be rather different. As was
mentioned already, observations of moving ejecta in XRB are most
likely related to relativistic motions detected in hotspots of radio
lobes in AGN \citep{lobanov_fomalont2001}.

\subsection{Temporal scales}

Flares and component ejections in XRB are often compared to accretion
disk instabilities in AGN, while the XRB state transition are commonly
related to grand scale accretion episodes in AGN
\citep{lobanov_celotti2005,lobanov_markoff2006,lobanov_belloni2006,lobanov_belloni2010,lobanov_marecki2010}.

With the simpleminded mass-scaling of timescales in XRB and AGN, these
comparisons face strong difficulties however. Flares and ejections of
new jet components in AGN may be related to characteristic instability
timescales in the disk at $20$--$200\,R_\mathrm{g}$
\citep{lobanov_lobanov2005}, which then should correspond to periods
of $\sim 2$\,sec in a $10\,M_{\odot}$ XRB. This is much shorter than
the typical hour-long timescales of component ejections in XRB -- and
the latter would have to be scaled up to $\sim 1000$ years in a
$10^8\,M_{\odot}$ black hole. At the same time, the typical month-long
scale of state transition in XRB would correspond to a one million
year scale in an AGN, which is seemingly shorter than the timescales
of radio active state of AGN as inferred from the sizes of the radio
jets and lobes.

Thus, if one adheres to the mass-scaling laws, one is forced to
conclude that observations of AGN jets: {\em a)}~probe linear scales
that are currently inaccessible for observations in XRB, and {\em
b)}~ejections in AGN and ejections in XRB may differ substantially in
their physics, with the latter behaving more like major AGN outbursts
lasting for decades.

Alternatively, one might suggest that the mass-scaling alone is not
sufficient and that environmental differences and energy-loss scales
(which are fundamentally similar in AGN and XRB) must be introduced in
the comparison as well. The energy losses can be a dominant factor in
determining timescales of evolution of flaring emission
\citep{lobanov_marscher1983,lobanov_marscher1985}, thus rending
practically irrelevant any comparison based on a simple ratio between
two emitting regions. Comparisons of variability timescales derived
from observations in different bands may be further complicated by
energy dependence of the emission losses. For instance, these timescales would
differ pretty much the same eight orders of magnitude for synchrotron
loss timescale $t_\mathrm{s} \propto B^2 E^1$, if X-ray variability in
XRB is compared to radio variability in AGN (not to forget about the
influence of the likely differences in the magnetic field strength in
XRB and AGN).

All these examples are only underlying the importance of continuing to
analyse carefully the similarities an differences between the
observational manifestations in AGN and XRB jets, and attempting to
provide a more physically complete basis for such comparisons.

\bibliographystyle{aa}

\begin{thebibliography}{}

\bibitem[Agudo et al.(2001)]{lobanov_agudo2001} Agudo, I., G\'omez,
J.L., Mart\'i, J.-M., et al., 2001, ApJ, 549, 183

\bibitem[Alberdi et al.(2000)]{lobanov_alberdi2000} Alberdi, A.,
G\'omez, J.L., Marcaide, J.M. et al.  2000, A\&A, 361, 529

\bibitem[Aloy et al.(2001)]{lobanov_aloy2001} Aloy, M.-A., Gomez, J.L.,
Iba\~nez, J.-M. et al. 2001, ApJ, 528, 85

\bibitem[Arshakian et al.(2010)]{lobanov_arshakian2010} Arshakian,
T.G., Le\'on-Tavares, J., Lobanov, A.P., et al. 2010, MNRAS, 401, 1231

\bibitem[Bach et al.(2005)]{lobanov_bach2005} Bach, U., Kadler, M., Krichbaum,
  T.P., et al., in \textit{Future Directions in High Resolution
    Astronomy: The 10th Anniversary of the VLBA}, J. Romney, M. Reed
  (eds.), ASP Conf. Ser., v.\,340, p.\,30

\bibitem[Becker et al.(2008)]{lobanov_becker2008} Becker, P.A., Das,
S., Le, T. 2008, ApJ, 677, L93

\bibitem[Belloni(2006)]{lobanov_belloni2006} Belloni, T.M. 2006,
Adv. Sp. Sci., 38, 2801

\bibitem[Belloni(2010)]{lobanov_belloni2010} Belloni, T.M. 2010,
Lecture Notes in Physics, 794, 53

\bibitem[Blandford \& Payne(1982)]{lobanov_blandford1982} Blandford,
R.D., Payne, D.G. 1982, MNRAS 199, 883

\bibitem[Blandford \& Znajek(1977)]{lobanov_blandford1977} Blandford,
R.D., Znajek, R.L. 1977, MNRAS, 179, 433

\bibitem[Bogovalov \& Tsynganos(2005)]{lobanov_bogovalov2005} Bogovalov,
S.V., Tsinganos, V. 2005, MNRAS 357, 918

\bibitem[Camenzind(2005)]{lobanov_camenzind2005} Camenzind, M. 2005,
MemSAIt, 76, 98

\bibitem[Celotti(2005)]{lobanov_celotti2005} Celotti, A. 2005, Ap\&SS,
230, 23

\bibitem[Celotti \& Fabian(1993)]{lobanov_celotti1993} Celotti, A., Fabian,
A.C. 1993, MNRAS 264, 228

\bibitem[Corbel et al.(2002)]{lobanov_corbel2002} Corbel, S., Fender,
R.P., Tzioumis, A.K. et al. 2002, Science, 298, 196

\bibitem[Daly \& Marscher(1988)]{lobanov_daly1988} Daly, R.A.,
Marscher, A.P. 1988, ApJ, 334, 539

\bibitem[D'Arcangelo et al.(2007)]{lobanov_darcangelo2007}
D'Arcangelo, F.D., Marscher, A.P., Jorstad, S.G., et al. 2007, ApJ,
659, L10

\bibitem[Dhawan et al.(2007)]{lobanov_dhawan2007} Dhawan, V., Mirabel,
I.F. Rib\'o, M., Rodrigues, I. 2007, ApJ, 668, 430

\bibitem[Falcke(2001)]{lobanov_falcke2001} Falcke, H. 2001, Rev. Mod. Astron.,
14, 15

\bibitem[Falcke et al.(2004)]{lobanov_falcke2004} Falcke, H., K\"ording,
E., Markoff, S., 2004, A\&A, 414, 895

\bibitem[Fender et al.(2004)]{lobanov_fender2004} Fender, R.P., Belloni,
T., Gallo, E. 2004, MNRAS, 355, 1105

\bibitem[Fomalont et al.(2001)]{lobanov_fomalont2001} Fomalont, E.B.,
Geldzahler, B.J., Bradshaw, C.F. 2001, AJ, 558, 283

\bibitem[Georganopoulos \& Kazanas(2004)]{lobanov_georganopoulos2004}
Georganopoulos, M., Kazanas, D. 2005, ApJ, 604, L81

\bibitem[G\'omez et al.(1995)]{lobanov_gomez1995} G\'omez, J.L.,
Marti, J.M., Marscher, A.P. et al. 1995, ApJ, 449, L19

\bibitem[G\'omez et al.(2001)]{lobanov_gomez2001} G\'omez, J.L.,
Marscher, A.P., Alberdi, A., et al. 2001, ApJ, 561, 161

\bibitem[G\'omez et al.(2008)]{lobanov_gomez2008} G\'omez J.L., Agudo, I.,
Marscher, A.P. et al. 2008, MemSAIt, 79, 1157

\bibitem[Gracia et al.(2005)]{lobanov_gracia2005} Gracia, J.,
Tsinganos, K., Bogovalov, S. V. 2005, A\&A, 442, 7

\bibitem[Hirotani(2005)]{lobanov_hirotani2005} Hirotani, K. 2005, 619, 73

\bibitem[Hirotani et al.(2000)]{lobanov_hirotani2000} Hirotani, K.,
Iguchi, S., Kimura, M, Wajima, K. 2000, ApJ, 545, 100

\bibitem[Hardee(2000)]{lobanov_hardee2000} Hardee, P.E. 2000, ApJ, 533, 176


\bibitem[Hardee(2007)]{lobanov_hardee2007} Hardee, P.E. 2007, ApJ, 664, 26

\bibitem[Hardee et al.(2005)]{lobanov_hardee2005} Hardee, P.E.,
Walker, R.C., G\'omez, J.L. 2005, ApJ, 620, 646

\bibitem[Hujeirat et al.(2003)]{lobanov_hujeirat2003} Hujeirat,
M. Livio, M. Camenzind,M., et al.\ 2003, A\&A 408, 415

\bibitem[Jorstad et al.(2005)]{lobanov_jorstad2005} Jorstad, S.G.,
Marscher, A.P., Lister, M.L., et al. 2005, AJ, 130, 1418

\bibitem[Junor(1999)]{lobanov_junor1999} Junor, W., Biretta, J.A.,
Livio, M.\ 1999, Nature 401, 891

\bibitem[Kadler et al.(2004)]{lobanov_kadler2004} Kadler, M., Ros. E.,
Lobanov, A.P., et al. 2004, A\&A, 426, 481

\bibitem[Kellermann et al.(2004)]{lobanov_kellermann2004} Kellermman,
K.I., Lister, M.L., Homan, D.C., et al. 2004, AJ, 609, 539

\bibitem[K\"ording et al.(2007)]{lobanov_koerding2007} K\"ording, E.G.,
Migliari, S., Fender, R. et al. 2007, MNRAS, 380, 301

\bibitem[K\"ording et al.(2008)]{lobanov_koerding2008} K\"ording, E.G.,
Jester, S., Fender, R. 2008, MNRAS, 383, 277

\bibitem[Koide et al.(2002)]{lobanov_koide2002} Koide, S., Shibata, K.,
Kudoh, T., et al.,\ 2002, Science 295, 1688

\bibitem[Komissarov(2005)]{lobanov_komissarov2005} Komissarov, S.S.\
2005, MNRAS 359, 801

\bibitem[Komissarov et al.(2007)]{lobanov_komissarov2007} Komissarov, S.S. 
Barkov, M.V., Vlahakis, N., K\"onigl, A. 2007, MNRAS, 380, 51

\bibitem[Krichbaum et al.(2008)]{lobanov_krichbaum2008} Krichbaum, T.P., Lee,
S.S., Lobanov, A.P., et al.,\ 2008, \textit{Extragalactic Jets: Theory
and Observation from Radio to Gamma Ray}, ASP Conf. Ser., v.\,386
eds. T.A. Rector \& D.S. De Young, (ASP: San Francisco), p.\,186

\bibitem[Lee et al.(2008)]{lobanov_lee2008} Lee, S.-S., Lobanov, A.P.,
Krichbaum, T.P., et al. 2008, AJ, 136, 159

\bibitem[Le\'on-Tavares et al.(2010)]{lobanov_leon2010}
Le\'on-Tavares, J., Lobanov, A.P., Chavushyan V.H., et al. 2010, ApJ,
715, 355


\bibitem[Lister et al.(2009)]{lobanov_lister2009} Lister, M.L., Cohen,
M.H., Homan, D.C., et al. 2009, AJ, 138, 1874

\bibitem[Lobanov(1998a)]{lobanov_lobanov1998a} Lobanov, A.P.\ 1998a, A\&A 390,
  79

\bibitem[Lobanov(1998b)]{lobanov_lobanov1998b} Lobanov, A.P.\ 1998b, A\&AS 132,
  261

\bibitem[Lobanov(2007)]{lobanov_lobanov2007} Lobanov, A.P.\ 2007, Ap\&SS, 311, 263

\bibitem[Lobanov(2008)]{lobanov_lobanov2008} Lobanov, A.P.\ 2008,
  Mem.S.A.It. 79, 1062

\bibitem[Lobanov(2010)]{lobanov_lobanov2010} Lobanov, A.P.\ 2010, "Fermi
meets Jansky - AGN in Radio and Gamma-Rays", Savolainen, T., Ros, E.,
Porcas, R.W., Zensus, J.A. (eds.), (MPIfR:Bonn), p.\,110, {\em arXiv:1010.2856} 

\bibitem[Lobanov \& Roland(2005)]{lobanov_lobanov2005} Lobanov, A.P.,
Roland, J.\ 2005, A\&A 431, 831

\bibitem[Lobanov \& Zensus(1999)]{lobanov_lobanov1999} Lobanov, A.P.,
Zensus, J.A. 1999, ApJ, 521, 590

\bibitem[Lobanov \& Zensus(2001)]{lobanov_lobanov2001} Lobanov, A.P.,
Zensus, J.A. 2001, Science, 294, 128

\bibitem[Lobanov \& Zensus(2006)]{lobanov_lobanov2006} Lobanov, A.P.,
Zensus, J.A.\ 2006, \textit{Exploring the Cosmic Frontier: Astrophysical
Instruments for the 21$^\mathrm{st}$ Century}, ESO Astrophysical
Symp. Series, ed. by A.P. Lobanov, J.A. Zensus, C. Cesarsky,
P.J. Diamond (Springer, Heidelberg) p.\,147, {\em astro-ph/0606143}

\bibitem[Lobanov et al.(1998)]{lobanov_lobanov1998c} Lobanov, A.P.,
Krichbaum, T.P., Wiztel, A., et al. 1998, A\&A, 340, 60

\bibitem[Lobanov et al.(2003)]{lobanov_lobanov2003} Lobanov, A.P.,
Hardee, P.E., Eilek, J.A. 2003, NewAR, 47, 629

\bibitem[Lobanov et al.(2006)]{lobanov_lobanov2006} Lobanov, A.P.,
Krichbaum, T.P., Witzel, A., Zensus, J.A. 2006, PASJ, 58, 253

\bibitem[Marecki \& Swoboda(2010)]{lobanov_marecki2010} Marecki, A.,
Swoboda, B. 2010, {\em arXiv:1010.651}

\bibitem[Markoff(2006)]{lobanov_markoff2006} Markoff, S. 2006, PoS(MQW6)035

\bibitem[Marscher(1983)]{lobanov_marscher1983} Marscher, A.P. 1983, ApJ, 264,
  296

\bibitem[Marscher \& Gear(1985)]{lobanov_marscher1985} Marscher, A.P.,
Gear, W.K. 1985, ApJ, 298, 114

\bibitem[Marscher et al.(2008)]{lobanov_marscher2008} Marscher, A.P.,
Jorstad, S.G., D'Arcangelo, F.D., et al. 2008, Nature, 452, 966

\bibitem[Mimica et al.(2007)]{lobanov_mimica2007} Mimica, P., Aloy
M.A., M\"uller, E. 2007, A\&A, 466, 93

\bibitem[Mimica et al.(2009)]{lobanov_mimica2009} Mimica, P., Aloy,
M.A., Agudo, I. et al. 2009, ApJ, 696, 1142

\bibitem[McClintock \& Remillard(2003)]{lobanov_mcclintock2003}
McClintock, J.E., Remillard, R.A. 2003, {\em astro-ph/0306213}

\bibitem[McHardy et al.(2006)]{lobanov_mchardy2006} McHardy, I.M.,
Koerding, E., Knigge, C. et al. 2006, Nature, 444, 730

\bibitem[Meier(1999)]{lobanov_meier1999} Meier, D.L.  1999 ApJ 522,
753

\bibitem[Meier(2001)]{lobanov_meier2001} Meier, D.L. 2001, Asrop. \&
Space Sci. Suppl., 276, 245

\bibitem[Meier(2003)]{lobanov_meier2003} Meier, D.L. 2003, New
Astron. Rev., 47, 667

\bibitem[Meier(2009)]{lobanov_meier2009} Meier, D.L. 2009, in
  \textit{Approaching Micro-Arcsecond Resolution with VSOP-2:
    Astrophysics and Technologies}, Y. Hagiwara, E. Fomalont,
  M. Tsuboi, Y. Murata (eds.), ASP Conf. Ser., v.\,402, p.\,342

\bibitem[Meier et al.(2001)]{lobanov_meier2001b} Meier, D.L., Koide, S.,
Uchida, Y. 2001, Science, 291, 84

\bibitem[Merloni \& Fabian(2002)]{lobanov_merloni2002} Merloni, A.,
Fabian, A.C.\ 2002, MNRAS 332, 165

\bibitem[Merloni et al.(2003)]{lobanov_merloni2003} Merloni, A., Heinz,
S., Di Matteo, T.  2003, MNRAS, 345, 1057


\bibitem[Middeleberg \& Bach(2008)]{lobanov_middelberg2008} Middelberg,
E., Bach, U. 2008, Reports on Progress in Physics, v.\,71, p.\,066901.

\bibitem[Mirabel(2010)]{lobanov_mirabel2010} Mirabel, I.F. 2010, {\em
these proceedings}

\bibitem[Mirabel \& Rodriguz(1994)]{lobanov_mirabel1994} Mirabel, I.F., Rodr\'iguez, L.F.
1994, Nature, 371, 46

\bibitem[Perucho \& Mart\'i(2007)]{lobanov_perucho2007a} Perucho, M.,
Mart\'i, J.M. 2007, MNRAS, 382, 526

\bibitem[Perucho et al.(2006)]{lobanov_perucho2006} Perucho, M.,
Lobanov, A.P., Mart\'i, J.-M., Hardee, P.E. 2006, A\&A, 456, 493

\bibitem[Perucho et al.(2007)]{lobanov_perucho2007} Perucho, M., Lobanov, A.P. 2007,
  A\&A 469, 23

\bibitem[Porth \& Fendt(2010)]{lobanov_porth2010} Port, O., Fendt,
C. 2010, ApJ, 709, 1100

\bibitem[Remillard \& McClintock(2006)]{lobanov_remillard2006} Remillard, R.A.,
McClintock, J.E. 2006, ARAA, 44, 49

\bibitem[Reynolds et al.(1996)]{lobanov_reynolds1996} Reynolds, C.S.,
Fabian, A.C., Celotti, A., Rees, M.J. 1996, MNRAS, 283, 873

\bibitem[Savolainen et al.(2006)]{lobanov_savolainen2006} Savolainen,
T., Wiik, K., Valtaoja, E., et al. 2006, ApJ, 647, 172

\bibitem[Semenov et al.(2004)]{lobanov_semenov2004} Semenov, V.,
Dyadechkin, S., Punsly, B. 2004, Science, 305, 978

\bibitem[Sikora et al.(2005)]{lobanov_sikora2005} Sikora, M., Begelman,
M.C., Madejski, G.M., et al.,\ 2005, ApJ 625, 72

\bibitem[Spruit et al.(1997)]{lobanov_spruit1997} Spruit, H.C.,
Foglizzo, T., Stehle, R. 1997, MNRAS 288, 333

\bibitem[Unwin et al.(1997)]{lobanov_unwin1997} Unwin, S.C., Wehrle,
A.E., Lobanov, A.P., et al. 1997, ApJ, 480, 596

\bibitem[Takahashi(2004)]{lobanov_takahashi2004} Takahashi, R.\ 2004,
ApJ, 611, 996

\bibitem[Vlahakis \& K\"onigl(2004)]{lobanov_vlahakis2004} Vlahakis, N.,
K\"onigl, A.  2004, ApJ, 605, 656

\bibitem[Walker et al.(2000)]{lobanov_walker2000} Walker, R.C., Dhawan,
V., Romney, J.D., et al.,\ 2000, ApJ 530, 233

\bibitem[Walker et al.(2001)]{lobanov_walker2001} Walker, R.C.,
Benson, J.M., Unwin, S.C., et al. 2001, ApJ, 556, 756

\bibitem[Wardle et al.(1998)]{lobanov_wardle1998} Wardle J.F.C., Homan,
D.C., Ojha, R., Roberts, D.H. 1998, Nature, 395, 457

\bibitem[Zensus(1997)]{lobanov_zensus1997} Zensus, J.A. 1997,
ARAA, 35, 607


\end{thebibliography}

\end{document}